\documentclass[preprint]{aastex}

\received{24 August 1999}
\accepted{11 December 1999}

\shorttitle{EUVE Observations of Virgo \& M87}
\shortauthors{Bergh\"ofer, Bowyer \& Korpela}

\begin{document}

\title{EUVE Observations of clusters of galaxies: Virgo and M87}

\author{Thomas W. Bergh\"ofer\altaffilmark{1,2} and Stuart
  Bowyer\altaffilmark{1} and Eric Korpela\altaffilmark{1}}

\affil{Space Sciences Laboratory, University of California, Berkeley,
        CA 94720-7450, USA}

\altaffiltext{2}{Hamburger Sternwarte, Universit\"at Hamburg, Gojenbergsweg 
112, D-21029 Hamburg, Germany}

\begin{abstract}
We present results of a re-analysis of archival EUVE data of the central
region of the Virgo cluster. We employ the new analysis method developed by
Bowyer, Bergh\"ofer \& Korpela (1999) and find diffuse emission reaching out
to a distance of $\approx 13\arcmin$. The spatial distribution of this flux
is incompatible with a thermal plasma origin. We investigate whether this 
emission is due to inverse Compton scattering of relativistic electrons against
the 3$^{\circ}$\,K black-body background. We show that this emission cannot be
produced by an extrapolation to lower energies of the observed synchrotron
radio emitting electrons and an additional component of low energy relativistic
electrons is required. In addition to discussing the diffuse EUV emission, we 
present an analysis of the EUV emission originating from the jet in M87.
\end{abstract}

\section{Introduction}

Observations with the {\it Extreme Ultraviolet Explorer} (EUVE) have provided
evidence that a number of clusters of galaxies produce intense EUV
emission (e.g., Bowyer et al. 1997). The initial explanation for this emission
was that it is produced by 
a diffuse, (5--10) $\times 10^5$K thermal gas component of the intracluster 
medium (ICM). Gas at these temperatures cools very rapidly, however, and there
is no obvious energy source to re-heat it (Fabian, 1996). Consequently, a 
number of other mechanisms have been investigated as the source of the 
emission.
Inverse Compton (IC) scattering of cosmic microwave background photons by
relativistic electrons present in the ICM was proposed as the source of the
observed EUV emission in the Coma cluster (Hwang 1997; En{\ss}lin \& Biermann
1998). However, Bowyer \& Bergh\"ofer (1998) have shown that the spatial
distribution of the EUV emission in this cluster is not consistent with IC
emission from the observed population of relativistic electrons. 

A variety of alternative explanations has been advanced
which dismiss the EUVE excess in clusters of galaxies. Most recently, 
Arabadjis \& Bregman (1999) argue that the EUV excess can be 
explained away by a different cross section for absorption by hydrogen and 
helium in the foreground ISM column. Bowyer, Bergh\"ofer \& Korpela (1999)
find that in some clusters this may account for some of the excess present
in the ROSAT PSPC data, however, this cannot explain the intense EUV
excesses found with EUVE.

Bowyer, Bergh\"ofer \& Korpela (1999) reexamined EUVE data 
of the clusters Abell 1795, Abell 2199, and Coma. They demonstrated that the 
initially reported results are based on an improper background subtraction. 
In previous analyses a flat background had been assumed. However, a detailed
investigation of blank field observations with the EUVE Deep Survey (DS) 
instrument shows that the background consists of two independent components, 
a non-photonic background 
and a background due to photons. 
The non-photonic background level can be determined in obscured regions of the 
detector and can be directly subtracted from the raw data. However, the 
photonic background is affected by telescope vignetting and must be treated 
differently. 

In the case of Abell 1795 and Abell 2199, Bowyer, Bergh\"ofer \& Korpela
(1999) show that the extent of the diffuse EUV emission is much smaller than 
earlier reported. Furthermore, the radial EUV emission profiles of these two 
clusters show a flux deficit compared to the soft energy 
tail of the X-ray emitting intracluster gas. These findings are consistent 
with the presence of strong cooling flows in Abell 1795 and Abell 2199.

In this paper we employ our new reduction method to EUVE archival data of the 
central part of the Virgo cluster. 
We compare our results with results derived from radio observations of this 
region. We consider the possibility that the observed diffuse EUV 
excess emission is due to an inverse Compton process of the known population 
of relativistic electrons in the ICM near M87. Furthermore, we investigate
the emission originating from the jet in M87 and compare our results with
observations at other wavelengths.

\section{Data and Data Analysis}  

The Virgo cluster has been observed for 36,000 s with the Deep Survey (DS)
Telescope of EUVE (Bowyer \& Malina, 1991). Data reduction was carried out
with the EUVE package built in IRAF which is especially designed to process
EUVE data. 

In order to reduce the non-photonic (non-astronomical) background contribution
to the data we investigated the pulse-height spectrum of all
detected events. 
A large number of EUVE DS observations of all kinds of astronomical
targets has shown that a typical pulse-height spectrum consists of two
components, a Gaussian profile representing the source events and an 
exponential background distribution. More details about the
different background contributions to the DS data and the method of
pulse-height thresholding can be found in Bergh\"ofer et al. (1998). 
From our experience with stellar and extragalactic observations with EUVE we 
know that the pulse-height selection effectively reduces the non-astronomical 
background in the data without any significant reduction of the source signal.
By comparing the source signal with and without pulse-height selection we find
that the effect on the source signal is lower than 4\%.

Then we applied corrections for detector dead time and for telemetry 
limitations (primbsching) to the
screened event list and produced a DS EUV image of the Virgo cluster.
We then determined the non-photonic background level in the image
from highly obscured regions at the outer most parts of the field of view near 
the Lexan/B filter frame bars. This non-astronomical background contribution 
is assumed to be constant over the entire detector field and was
subtracted from the image.

In order to subtract the (vignetted) photonic background we computed the 
azimuthally averaged radial emission profile centered on M87. We used the EUVE
DS sensitivity map provided by Bowyer, Bergh\"ofer \& Korpela (1999) to 
determine a radial sensitivity profile centered on the detector position of
M87. This was then fit to the outer part (15--20\arcmin) of the radial emission
profile to determine the scaling factor between sensitivity profile and the 
photonic background in the data. The radial emission profile and the best fit 
background model are shown in Figure\ \ref{m87rad}.

\section{Results}
\label{results}

The data in Figure\ \ref{m87rad} demonstrate the presence of diffuse EUV emission in the 
vicinity of M87 which extends to 
a radius of $\approx$13\arcmin. At larger radii the radial profile is well fit
by the background model demonstrating the absence of any significant 
cluster emission beyond this. The initial publication on the diffuse EUV emission from
Virgo (Lieu et al. 1996) claimed to detect excess emission to 20\arcmin.

In Figure\ \ref{m87sig} we plot the background subtracted radial EUV emission
profile (solid line). The dashed line shows the expected EUV emission of the
low energy tail of the X-ray emitting diffuse intracluster gas as derived
in the following.
Note that the inner 1\arcmin\ bin is dominated by the core and jet of M87 and 
must be ignored for the discussion of the diffuse emission.

To determine the diffuse X-ray contribution to the observed EUV emission we
processed ROSAT PSPC archival data of the Virgo cluster. We used standard 
procedures implemented in the EXSAS software package to produce an image from 
the photon event list. Then a vignetting corrected exposure map was  
computed for this data set and a PSPC count rate image was generated by 
dividing the PSPC image by the exposure map.

We point out that the background in the ROSAT PSPC hard energy band is 
dominated by the photonic (vignetted) background and the contribution of 
the non-photonic background is minor. Therefore, a similar analysis as 
described for the EUVE DS data including a separation of the photonic and 
non-photonic background contributions is not essential. However, in the case of
detectors with low effective areas (e.g., BeppoSAX), and less efficient 
rejection mechanisms for non-photonic events, this background contribution must
be treated separately.

For our analysis of the ROSAT PSPC data we selected only photon events in the 
hardest energy band, channels 90--236. This channel selection has several 
advantages: First, any contamination by a possible steep-spectrum source at soft
X-ray energies is excluded and, therefore, ensures that this band pass 
represents only thermal contributions to the overall diffuse emission in Virgo.
Second, this part of the ROSAT band pass is nearly unaffected by interstellar
absorption. This minimizes errors due to possible differential ISM absorption
effects when modeling conversion factors between DS and PSPC 
counts. Third, the count rate conversion factor between DS and PSPC is nearly
temperature independent in the range of X-ray temperatures measured in the
central Virgo region and, thus, ROSAT count rates of the diffuse X-ray
emission can be converted into DS count rates by using one single conversion
factor.

In order to be able to convert PSPC counts into DS counts we modeled
conversion factors for a range of plasma temperatures (0.1--2.7 keV) employing
the MEKAL plasma emission code with abundances of 0.34 solar (Hwang et
al. 1997). These calculations include absorption by the interstellar
medium. We used an ISM absorption column density of $1.72 \times
10^{20}$cm$^{-2}$ (Hartmann \& Burton 1997) and an absorption model including 
cross sections and ionization ratios for the ISM as described in Bowyer, 
Bergh\"ofer \& Korpela\ (1999). In Figure\ \ref{m87theo} we show the 
DS to PSPC count rate conversion factor. The left-hand scale and the solid
curve gives the plasma temperature as a function of the DS to PSPC count rate 
ratio. As can be seen, for a wide range of temperatures
(0.6--2.7 keV) the model conversion factor is constant within 15\%. According
to B\"ohringer et al.\ (1995) and B\"ohringer (1999), the temperature
of the X-ray emitting
intracluster gas in the Virgo cluster is $\approx$2 keV. In addition to this
thermal gas component these authors detected several diffuse emission features
near M87 which are significantly softer than the average Virgo cluster gas
temperature. However, spectral fits to the ROSAT data do not provide any 
evidence for gas at
temperatures below 1 keV (B\"ohringer, private communication). For
temperatures near 1 keV the modeled conversion factor for a thermal gas is
slightly lower than for higher temperatures. Therefore, the contribution of 
the lower temperature components to the overall diffuse X-ray emission in 
the EUV band pass is lower than the dominant 2 keV cluster gas 
component. Using the conversion factor appropriate for the mean cluster gas 
temperature of 2 keV for the entire emission including the softer thermal 
enhancements, slightly overestimates the low energy X-ray contribution to the 
EUV emission.

We also modeled the DS to PSPC conversion factor
for a non-thermal power law type spectrum including ISM absorption. The 
right-hand scale and dashed curve give the power law spectral index as a 
function of the modeled conversion factor.

In Figure\ \ref{m87ratio} we show the observed ratio between azimuthally 
averaged radial intensity profiles observed with the EUVE DS and 
PSPC. Within the error bars the ratio is constant (reduced $\chi^2$ = 0.9). 
The best fit value is $0.0186 \pm 0.0057$. The ratio for the inner 1\arcmin\ 
bin is consistent with this value, however, we excluded this point due to the 
presence of 
emission from the core and jet of M87. Sarazin \& Lieu (1998) have suggested
that an increasing EUV to X-ray emission ratio towards larger distances from 
the cluster center is an indication of an inverse Compton process producing
the EUV emission in the cluster. However, the data in Figure\ \ref{m87ratio} 
demonstrate that this is not observed in the central Virgo region.

Our best fit value of 0.0186 is $\approx$4.3 times larger
than expected for the low energy tail of the X-ray emitting gas in the Virgo 
cluster. Therefore, the X-ray contribution to the observed EUV excess in the 
central part of the Virgo cluster must be minor. 

It is clear that the ratio between observed EUV flux and modeled X-ray 
contribution cannot directly be used to determine the physical
parameters of the source. Instead, one must first subtract the X-ray 
contribution from the observed EUV emission. 

In Figure\ \ref{imaeuv} we show the spatial distribution of the EUV excess 
emission in the central Virgo region; the background and the contribution of
the low energy tail of the X-ray emitting ICM have been subtracted.
The central emission peak at the position of M87 is surrounded by
a diffuse EUV emission structure which is asymmetric in shape. Its extent
varies between 1\arcmin\ and 7\arcmin.
Several arm-like features are visible. At larger radii the EUV
emission results from a number of apparently discrete and extended diffuse 
features in the M87 radio halo region. These emission features are consistent 
with the emission seen in the surface brightness profile 
(Figure\ \ref{m87sig}) between 9--13\arcmin. These asymmetric features show the
flux is not produced by a gravitationally bound thermal gas. 
For the diffuse EUV emission within 7\arcmin\ (excluding the core + jet 
emission in the inner 1\arcmin) we determine a total count rate of 
$(0.036 \pm 0.006)$ counts\,s$^{-1}$. Assuming an extraction radius of 
13\arcmin\ results in a total count rate of $(0.066 \pm 0.009)$ 
counts\,s$^{-1}$.

We also investigated the EUV emission peak at the position of M87. X-ray 
observations with the {\em Einstein} and ROSAT HRIs have demonstrated that the 
central X-ray emission peak splits into two major components which are 
associated with the core and mainly knots A+B+C of the jet in M87.
The spatial resolution of the EUVE DS ($\approx$ 20\arcsec) is not sufficient 
to completely resolve the jet from the galaxy core. However, the central peak 
indicates emission slightly elongated by about one resolution element in the 
direction from the core to the jet. 
The central emission peak (core + jet) provides a total count rate 
of $(4.9 \pm 0.6) \times 10^{-3}$ counts\,s$^{-1}$ in excess of the diffuse 
emission component.

\section{Discussion and Conclusions}
\label{discuss}

\subsection{Diffuse EUV emission}
\label{disdiffuse}

The results of our reanalysis show a clear EUV excess in the central Virgo
region around M87. Compared to previous studies the azimuthally averaged
extent of this emission
is smaller and extends only to $\approx$13\arcmin\ from the center of M87.

To explore the nature of the EUV excess we compare this
emission with a 90 cm radio map of the central Virgo region near M87 (Owen,
Eilek \& Kassim 1999). If the diffuse EUV emission is due to inverse Compton
processes in the ICM, one would expect to see similar emission features in both
the EUV and radio image. In Figure\ \ref{radeuv} we show a contour plot of the
EUV emission superposed on the 90 cm radio map. As can be seen, the EUV
emission peaks at the position of the radio emission of the core and jet of
M87. EUV excess emission features are, however, not directly coincident with
any of the other brighter features visible in the radio map. 
The EUV emission is also not associated with the higher temperature
X-ray emission features seen in the ROSAT PSPC images in 
Virgo (cf. B\"ohringer 1999 and Harris 1999).

We next investigate whether the integrated flux of the diffuse EUV emission
is compatible with an inverse Compton origin of the observed EUV excess in
the central Virgo region. We use the observed radio synchrotron power 
law spectrum of the M87 halo ($\alpha = 0.84$, Herbig \& Readhead 1992) to
compute the underlying 
distribution of relativistic electrons in this region and its inverse Compton
flux. Note that the radio spectrum needs to be extrapolated into the low
frequency range near 1 MHz which is not observable due to ionospheric effects.
The conversion from the synchrotron spectrum into an electron energy
distribution depends on the magnetic field strength in the ICM.
We derive a relation between magnetic field strength and the inverse
Compton flux produced by the relativistic electrons; the results are shown
in Figure\ \ref{icrates}.
The flux is folded with the EUVE DS response and given in
units of DS counts\,s$^{-1}$ which allows a direct comparison to the observed 
integrated DS count rate of the diffuse emission (horizontal line in Figure\
\ref{icrates}). As can be seen, for a magnetic field strength of 
$\approx 3\mu$G the observed flux matches the model flux. Note that
this value would also be consistent with Faraday rotation measurements in the
M87 halo (Dennison 1980).

However, with $\alpha = 0.84$ the radio synchrotron spectrum is inconsistent
with the required steep EUV to X-ray power law spectrum. In Figure\ 
\ref{m87theo} we show three dotted vertical lines labeled with 100\%, 10\%,
and 5\%. These lines indicate relative contributions of the hard energy tail 
of the EUV excess component to the overall X-ray emission in the 
ROSAT band. A contribution of 100\% is obviously not realistic since this 
would require that no emission is seen from the gravitationally bound 
intracluster gas. The other two dotted lines show 10\% and 5\% contributions, 
respectively. No other emission component in excess of the thermal
component has been detected in the ROSAT PSPC data of Virgo and only
an upper limit
can be derived from this data. A determination of an accurate upper limit for
the EUV excess component in the ROSAT band is highly model dependent. However,
from our experience with ROSAT data of diffuse sources we estimate that a 
contribution of 10\% should be detectable. If we assume a 10\% contribution
as the upper limit for the EUV excess component in the ROSAT band, 
according to Figure\ \ref{m87theo} a power law photon number index of $\alpha 
\geq 3.2$ is required to explain the observed EUV flux and the upper limit in 
the ROSAT PSPC hard band (channels 90--238) by a non-thermal power law
source. Therefore, inverse Compton emission from the known 
population of relativistic electrons in the M87 halo cannot account for the
observed EUV excess in the central Virgo region.

We compute the total luminosity of the diffuse EUV emission for a steep 
non-thermal power law spectrum and for a low temperature thermal plasma 
spectrum since these have been discussed in the literature, but we make no 
claim that either of these are the correct spectral distribution for the
emission. 
Assuming a power law spectrum with $\alpha = 3.2$ results in a luminosity of
$5.2 \times 10^{42}$ erg\,s$^{-1}$ in the 0.05--0.2 keV band.
For a thermal plasma with a temperature of 0.15 keV we obtain a luminosity of
$5.7 \times 10^{42}$ erg\,s$^{-1}$. These values were derived from the total 
count rate of the diffuse EUV emission within 7\arcmin. Including the 
apparently discrete and extended diffuse EUV emission detected between 
7\arcmin\ and 13\arcmin\ increases the luminosity by 80\%. Assuming larger 
power law indices or lower plasma temperatures result in higher luminosities.
For the luminosity calculations we assume a distance of 17\,Mpc.

\subsection{EUV emission of the jet in M87}

Since the core and jet of M87 cannot be resolved in the EUVE image of M87, we
assume that the X-ray flux ratio between core and jet which can be determined 
from the ROSAT HRI observations is also valid for the EUV fluxes. Harris, 
Biretta \& Junor (1997) give a ratio of $\approx$ 1.5 for the core/jet X-ray flux 
ratio. Based on their compilation of measurements for the jet in M87,
Meisenheimer et al. (1996) derived a spectral index of 0.65 for the radio to
near-UV spectrum. In order to be able to explain the X-ray emission of the jet
in M87 by the same spectrum, these authors introduced a spectral cut-off near
10$^{15}$Hz. The spectral index of the UV to X-ray power law spectrum then has
to be $\alpha \approx 1.4$ to explain the UV and X-ray data.

Based on these assumptions we compute a flux of 
$3.4 \times 10^{-12}$\,erg\,cm$^{-2}$\,s$^{-1}$ ($6.5 \times 10^{-6}$\,Jy) 
and a luminosity of $1.2 \times 10^{41}$\,erg\,s$^{-1}$ for the emission of 
the M87 jet in the EUVE DS bandpass. For the luminosity calculation we assume 
a distance of 17\,Mpc.

In Figure\ \ref{jetspec} we show the radio-to-X-ray spectrum of the jet in M87
including the EUVE data point. As can be seen, the spectral model provided by 
Meisenheimer et al. (1996) also fits the EUVE observations. This confirms the 
suggested cut-off in the UV and further supports that the entire jet emission,
from the radio to the X-ray band, is synchrotron radiation produced by 
relativistic electrons in the jet.

\section{Summary}

The observed EUV excess in the central Virgo region is not spatially
coincident with either the distribution of the radio emission or the
observed high temperature thermal X-ray emission seen in the ROSAT
images. This provides strong evidence that a separate source mechanism
is present. In addition, due to the required steep EUV to X-ray spectrum, 
this emission cannot be produced by an extrapolation to lower energies of the 
observed synchrotron radio emitting electrons. If the observed EUV 
excess is inverse Compton emission, a new population population of 
relativistic electrons is required. Therefore, the same difficulties as in the
case of the explanation of the EUV excess of the Coma cluster (cf. Bowyer \& 
Bergh\"ofer 1998) exist in the central Virgo region.
The EUVE observations of M87 are consistent with the 
spectral cut-off in the spectrum of the jet in M87 as suggested by 
Meisenheimer et al. (1996). This further supports the idea that the EUV and
X-ray emission of the jet is synchrotron radiation.

\acknowledgments

We thank Jean Eilek for providing us a postscript file of the M87 radio map.
We acknowledge useful discussions with John Vallerga, Jean Dupuis, and Hans
B\"ohringer. This
work was supported in part by NASA contract NAS 5-30180. TWB was supported in 
part by a Feodor-Lynen Fellowship of the Alexander-von-Humboldt-Stiftung.


\figcaption[f1.eps] { \label{m87rad} The azimuthally averaged radial intensity
  profile of the EUV emission in the central part of Virgo (centered on M87)
  is shown as a solid line. The dashed line is the vignetted background. There
  is no EUV emission beyond $\approx$13\arcmin.}

\figcaption[f2.eps] { \label{m87sig} The solid line shows the azimuthally
  averaged radial intensity profile of the background subtracted EUV emission
  in the central part of Virgo. The dashed line provides the contribution of
  the low energy tail of the X-ray emitting intracluster gas. The central
  1\arcmin\ bin is dominated by the core and jet in M87 and must be ignored in
  the context of diffuse emission.}

\figcaption[f3.eps] { \label{m87theo} Conversion factors from ROSAT PSPC count
  rates into EUVE DS rates modeled for a spectrum of a thermal plasma model and
  a non-thermal power law type source including the ISM absorption towards the
  Virgo cluster. Ordinates: Left-hand scale and solid curve give the plasma 
  temperature for the thermal model; right-hand scale and dashed curve give the
  power law spectral index for the non-thermal model. Three dotted vertical 
  lines labeled with 100\%, 10\%, and 5\%. These lines indicate relative 
  contributions of the hard energy tail of the EUV excess component to the 
  overall X-ray emission in the analyzed ROSAT band (see Sect.\ 
  \ref{disdiffuse}).}

\figcaption[f4.eps] { \label{m87ratio}  The ratio between observed azimuthally
 averaged intensity profiles observed with the EUVE DS and ROSAT PSPC. The
 dotted line represents the best fit value of 0.186 assuming a constant ratio.}

\figcaption[f5.eps] { \label{imaeuv} Spatial distribution of the EUV excess
  emission in the central Virgo region. The contribution of the low energy
  tail of the X-ray emitting ICM has been subtracted from the data.}

\figcaption[f6.eps] { \label{radeuv} Contour plot of the background and X-ray
  subtracted EUV emission shown superposed on the 90 cm radio map of the M87
  halo. The contours provide fluxes of 1.5, 4.5, 7, 10, 13, and 27 $\sigma$
  above the noise level.}

\figcaption[f7.eps] { \label{icrates} EUVE DS count rate as a function of
  magnetic field strength. The horizontal line provides the observed total
  count rate for the diffuse emission component. The error in this value is
  shown as a gray shaded area.}

\figcaption[f8.eps] { \label{jetspec} Radio-to-X-ray spectrum of the jet in M87
 The solid line shows the model provided by Meisenheimer et al. (1996). The 
 EUVE data point is well fit by this model with a cut-off at UV wavelengths.}

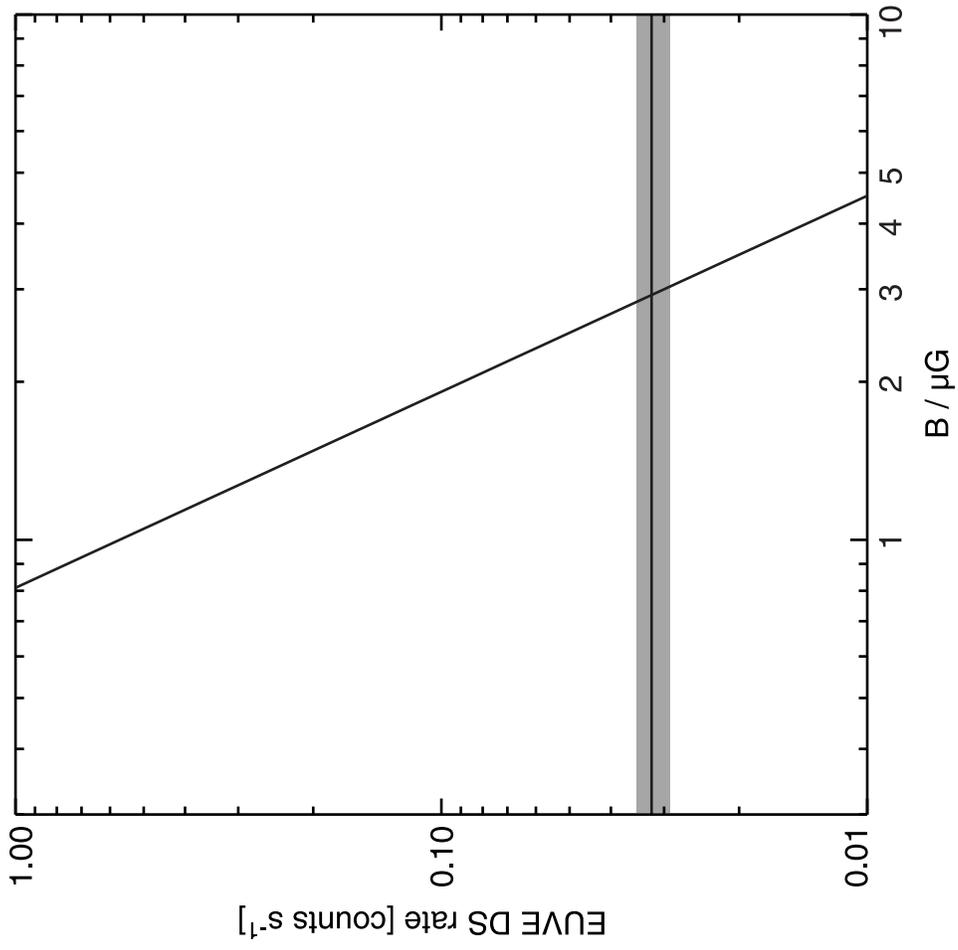
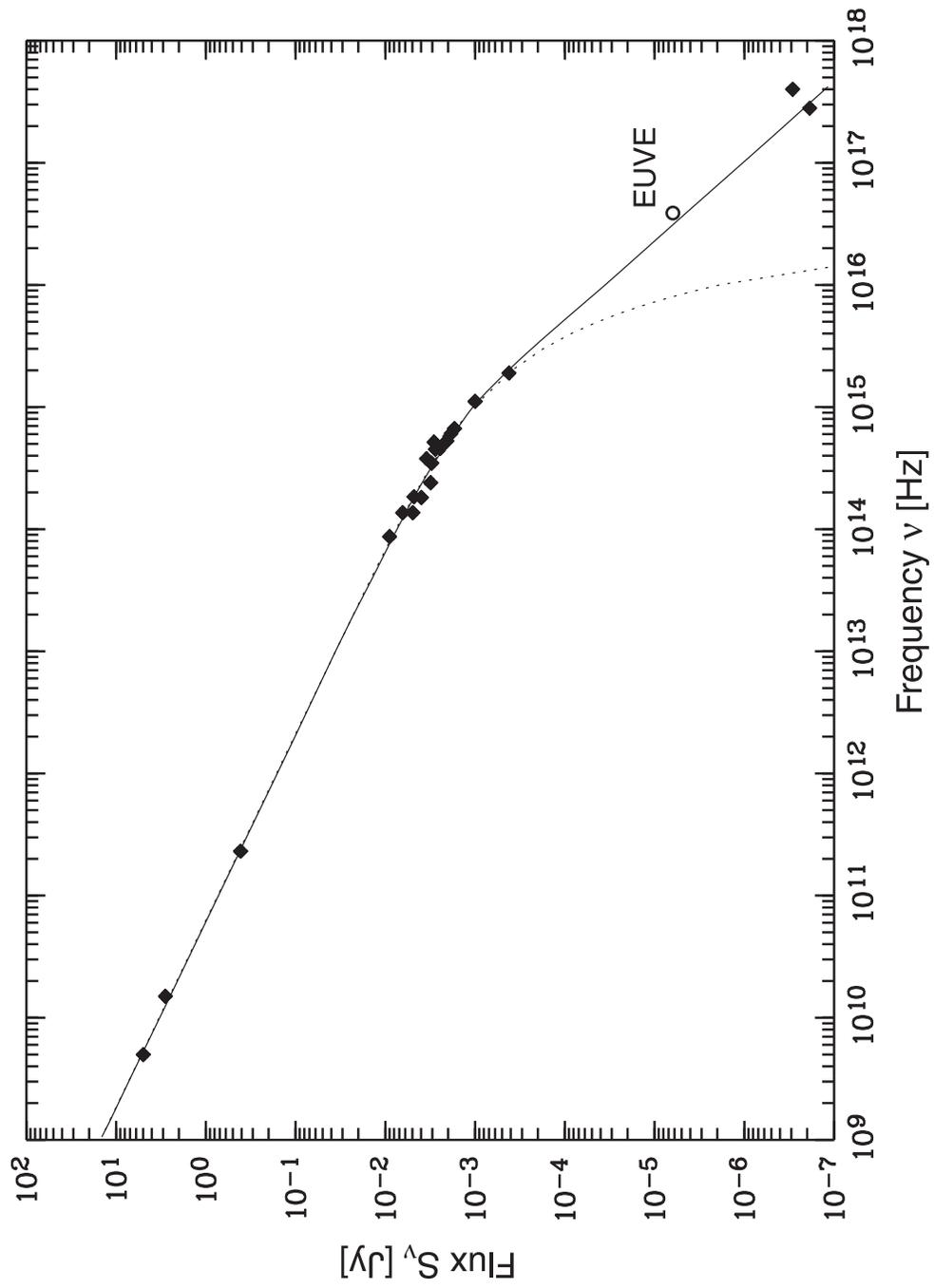
\end{document}